\documentclass{article}

\usepackage{arxiv}
\usepackage{multirow}

\usepackage{amsmath, amsthm, amssymb, amsfonts}
\usepackage{enumitem}

\usepackage[numbers,square]{natbib}

\usepackage[utf8]{inputenc}	
\usepackage[T1]{fontenc}	
\usepackage{xcolor}		
\usepackage[colorlinks = true,
            linkcolor = purple,
            urlcolor  = blue,
            citecolor = cyan,
            anchorcolor = black]{hyperref}	
\usepackage{booktabs} 		
\usepackage{nicefrac}		
\usepackage{microtype}		
\usepackage{float}			
\usepackage{graphicx}        
\usepackage{lineno}		    
\usepackage{lipsum}		    
\usepackage{doi}            

\usepackage{newfloat}
\DeclareFloatingEnvironment[name={Supplementary Figure}]{suppfigure}
\usepackage{sidecap}
\sidecaptionvpos{figure}{c}

\usepackage{titlesec}
\titlespacing\section{0pt}{12pt plus 3pt minus 3pt}{1pt plus 1pt minus 1pt}
\titlespacing\subsection{0pt}{10pt plus 3pt minus 3pt}{1pt plus 1pt minus 1pt}
\titlespacing\subsubsection{0pt}{8pt plus 3pt minus 3pt}{1pt plus 1pt minus 1pt}

\usepackage{tikz,xcolor,hyperref}
\definecolor{lime}{HTML}{A6CE39}
\DeclareRobustCommand{\orcidicon}{
	\begin{tikzpicture}
	\draw[lime, fill=lime] (0,0)
	circle [radius=0.16]
	node[white] {{\fontfamily{qag}\selectfont \tiny ID}};
	\draw[white, fill=white] (-0.0625,0.095)
	circle [radius=0.007];
	\end{tikzpicture}
	\hspace{-2mm}
}
\foreach \x in {A, ..., Z}{\expandafter\xdef\csname orcid\x\endcsname{\noexpand\href{https://orcid.org/\csname orcidauthor\x\endcsname}
			{\noexpand\orcidicon}}
}

\title{Mining Asymmetric Intertextuality}

\newif\ifuniqueAffiliation
\uniqueAffiliationfalse 

\author{
  Pak Kin Lau\orcidA{} \\
  Department of Mathematics \\
  Department of History \\
  The Chinese University of Hong Kong \\
  \texttt{pakkinlau@cuhk.edu.hk} \\
  \And
  Stuart Michael McManus\orcidB{} \\
  Department of History \\
  The Chinese University of Hong Kong \\
  \texttt{smcmanus@cuhk.edu.hk} \\
}

\hypersetup{
pdftitle={Mining Asymmetric Intertextuality},
pdfsubject={cs.CL, cs.IR},
pdfauthor={Pak Kin Lau, Stuart Michael McManus},
pdfkeywords={Asymmetric Intertextuality, Metadata, Hierarchical Chunking, LLM, Split-Normalize-Merge, Vector Search},
}

\begin{document}

\date{October 16, 2024}

\maketitle

\begin{abstract}
This paper introduces a new task in Natural Language Processing (NLP) and Digital Humanities (DH): Mining Asymmetric Intertextuality. Asymmetric intertextuality refers to one-sided relationships between texts, where one text cites, quotes, or borrows from another without reciprocation. These relationships are common in literature and historical texts, where a later work references aclassical or older text that remain static.

We propose a scalable and adaptive approach for mining asymmetric intertextuality, leveraging a split-normalize-merge paradigm. In this approach, documents are split into smaller chunks, normalized into structured data using LLM-assisted metadata extraction, and merged during querying to detect both explicit and implicit intertextual relationships. Our system handles intertextuality at various levels, from direct quotations to paraphrasing and cross-document influence, using a combination of metadata filtering, vector similarity search, and LLM-based verification.

This method is particularly well-suited for dynamically growing corpora, such as expanding literary archives or historical databases. By enabling the continuous integration of new documents, the system can scale efficiently, making it highly valuable for digital humanities practitioners in literacy studies, historical research and related fields.

\end{abstract}


\vspace{0.35cm}

\section{Introduction}

As the volume of literary, and historical texts available in digital form continues to grow exponentially, the ability to trace how ideas are referenced, borrowed, or transformed across texts has become ever more vital. \textit{Intertextuality}—the study of relationships between texts—provides critical insight into intellectual traditions, cultural change, and literary influence. However, the increasing complexity of digital archives poses significant challenges for tracking these relationships, particularly when they are subtle or implicit.

One particularly challenging form of intertextuality is \textit{asymmetric intertextuality}. In this paper, we define asymmetric intertextuality as a type of relationship between two texts in which they are structurally or semantically \textit{non-isomorphic}—that is, they do not exhibit a straightforward one-to-one correspondence in their relationship. This occurs when one text references or borrows from another, but the connection between them is non-reciprocal and structurally uneven. While \textit{one-way intertextuality} refers to a directional flow of influence, \textit{asymmetric intertextuality} encompasses a broader mismatch in how the texts relate, where the influence cannot be mapped symmetrically or bidirectionally \citep{genette1997palimpsests}.

In contexts such as literature (allusions and thematic borrowing), journalism (quotations), and historical texts (quotations, allusions, borrowings, etc), both \textit{one-way} and \textit{asymmetric} intertextual relationships often occur. Despite concerted efforts to build tools and develop techniques to identify intertextuality, these relationships are not easily detected by traditional methods, including token-based n-gram analysis or even recent advancements in semantic search. The detection of such asymmetric relationships requires a more advanced system capable of identifying both explicit and implicit connections, especially when texts transform or paraphrase the original content \citep{allen2000intertextuality}.

In this paper, we propose to tackle the problem of detecting \textit{asymmetric intertextuality} using a novel \textit{Split-Normalize-Merge paradigm}. Our approach breaks documents into smaller chunks (\textit{Split}), normalizes them using LLM-assisted metadata extraction (\textit{Normalize}), and reassembles them during querying to detect both explicit and implicit relationships (\textit{Merge}). By combining large language models (LLMs), vector similarity search, and metadata filtering, we address the limitations of traditional approaches and propose a scalable solution for detecting asymmetric intertextuality across large and dynamically growing corpora.

\subsection{Broader Definition of Intertextuality}

Intertextuality encompasses a wide range of textual relationships, from direct quotations to more subtle forms of influence like allusion, plagiarism, or thematic borrowing. Scholars have long studied how texts reference one another, but the complexity of intertextuality increases significantly when references are implicit or indirect.

For the purposes of this paper, we focus on \textit{explicit} forms of intertextuality, such as \textit{quotation} and \textit{citation}, but we also acknowledge the broader landscape that includes more implicit forms like allusion, paraphrasing, and thematic borrowing.

\begin{figure}[H]
    \centering
    \includegraphics[width=0.7\columnwidth]{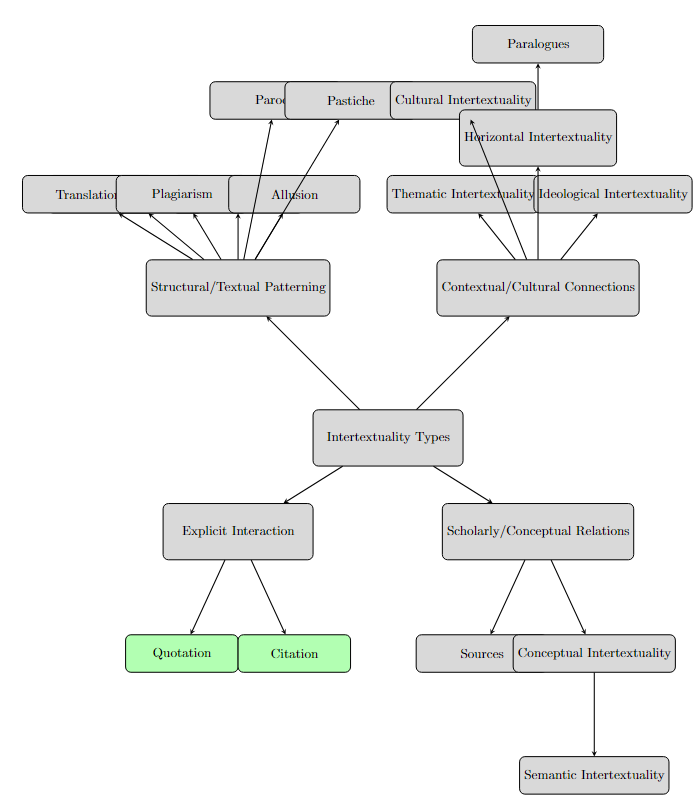}
    \caption{Types of intertextuality. Where Quotation and Citation (colored in green) has obvious cues.}
    \label{fig:intertext_types}
\end{figure}

Forms of intertextuality can include:

\begin{enumerate}
    \item \textbf{Explicit Interaction} involves direct textual references or borrowings:
    \begin{itemize}
        \item \textbf{Quotation}: The verbatim or near-verbatim reuse of text. For example, Herman Melville's \textit{Moby Dick} directly quotes biblical passages like the story of Jonah to draw theological parallels \citep{melville1851moby}.
        \item \textbf{Citation}: While less common in historical and literary texts, citation-like references still appear, such as in Chinese classical texts that reference earlier works \citep{sturgeon2018digital}.
    \end{itemize}

    \item \textbf{Scholarly/Conceptual Relations} focus on intellectual and thematic connections:
    \begin{itemize}
        \item \textbf{Sources (Implicit citation)}: The identification of original works that provide foundational content for new texts \citep{romanello2016sources}.
        \item \textbf{Conceptual Intertextuality}: The sharing of themes or ideas across texts without direct textual borrowing. For instance, classical Chinese texts often reference philosophical concepts from earlier works without direct quotation \citep{vierthaler2019blast}.
    \end{itemize}
    
    \item \textbf{Structural/Textual Patterning} refers to intertextuality driven by structural or stylistic similarities:
    \begin{itemize}
        \item \textbf{Translation}: The transformation of a text from one language to another \citep{bahdanau2014neural}.
        \item \textbf{Plagiarism}: Unauthorized textual overlap \citep{ghiban2013plagiarism}.
        \item \textbf{Allusion}: Indirect references to earlier works, relying on the reader's knowledge \citep{henrichs2020allusions}.
        \item \textbf{Paraphrastic Intertextuality}: Rephrasing content while maintaining semantic similarity \citep{newell2018automatic}.
        \item \textbf{Parody and Pastiche}: The imitation of a text with satirical or stylistic intent \citep{miola2004intertextuality,kabbara2016stylistic}.
    \end{itemize}

    \item \textbf{Contextual/Cultural Connections} explore intertextuality in relation to broader cultural or thematic frameworks:
    \begin{itemize}
        \item \textbf{Thematic Intertextuality}: For example, classical Chinese texts engage in thematic intertextuality through philosophical ideas that transcend specific works \citep{tharsen2022textpair}.
    \end{itemize}
\end{enumerate}

In this paper, we focus specifically on quotation and citation, while acknowledging that asymmetry in intertextuality is often evident in less explicit forms, such as paraphrasing or thematic borrowing.

\subsection{One-Way Intertextuality: A Cross-Domain Phenomenon}

While \textit{asymmetric intertextuality} involves structural or semantic transformations, \textit{one-way intertextuality} refers more simply to the \textit{directional flow} of references, communication, or influence between two texts. This phenomenon is widespread across various domains, including literature, journalism, and historical texts.

In \textit{one-way intertextuality}, a newer text draws from or references an older one, but the reference is not reciprocated. This one-directional flow is common in many contexts and can take several forms, including \textit{quotation}, \textit{paraphrase}, and \textit{thematic borrowing} \citep{standage2013writing}.

\subsubsection{Examples Across Domains}

\begin{itemize}
    \item \textbf{Literature}: Many pre-modern novels or poems reference or allude to classical works (e.g., Shakespeare, Homer, or Chinese classics like the \textit{Shijing}). However, these allusions are one-sided: the original works remain static and unchanged, while the newer texts reinterpret or recontextualize them \citep{sturgeon2018digital}.
    
    \item \textbf{Journalism}: News articles frequently quote or paraphrase statements from public figures or historical documents. However, these quotes are often taken out of context or reframed to fit the narrative of the article, making the relationship non-reciprocal and often asymmetric.
\end{itemize}

\subsubsection{Challenges in Detecting One-Way Intertextuality}

Detecting \textit{one-way intertextuality} is challenging because the references are often implicit. Traditional \textit{token-based methods} (like keyword matching or n-gram analysis) struggle to capture paraphrased or recontextualized references, while \textit{semantic search methods} often miss subtle thematic or structural borrowings \citep{allen2000intertextuality}.

For example, a modern text may \textit{paraphrase} an older work without using direct quotations, making it hard for basic search algorithms to detect the connection. Similarly, thematic borrowings, where ideas or concepts are reinterpreted without explicit citation, pose a major challenge for current intertextuality detection tools.

\subsubsection{Why One-Way Intertextuality is Usually Asymmetric}

Not all one-way relationships are \textit{asymmetric}, but many are. In cases where a newer text transforms, paraphrases, or recontextualizes an older one, the relationship between the texts becomes structurally \textit{non-isomorphic}. The newer text may borrow heavily from the older one, but the transformation of the content introduces asymmetry. The two texts are no longer equivalent in their relationship, and the influence is not easily mapped in a bi-directional way.

This is particularly true in cases of \textit{paraphrasing} or \textit{thematic borrowing}, where the newer text changes the meaning or application of the original content. The lack of direct quotation or citation makes these connections harder to detect, requiring more advanced systems that can capture the \textit{semantic and structural differences} between the two texts.

\subsection{Common Characteristics and Challenges Across Domains}

While one-way and asymmetric intertextuality manifest differently across various domains, they share several common characteristics that make them difficult to detect using traditional methods:

\begin{itemize}
    \item \textbf{Unidirectional Influence}: Whether in academic writing or literature, the flow of influence is always one-way. A newer text references an older one, but the older text cannot reciprocate.
  
    \item \textbf{Implicitness}: Many one-way intertextual links are implicit. A modern text may refract, paraphrase, or borrow thematically from an older work without directly citing it or using recognizable keywords, making detection difficult.

    \item \textbf{Paraphrasing and Transformation}: One of the biggest challenges in detecting one-way intertextuality is that newer texts often transform or paraphrase the content of the older work. This transformation introduces asymmetry, posing challenges to traditional search methods.

    \item \textbf{Cross-Domain Applicability}: Whether in academia, literature, journalism, or social media, one-way and asymmetric intertextuality appear across a variety of domains, making the need for a cross-domain detection system more pressing.
\end{itemize}

\subsection{A New Formalism for Detecting Asymmetry in Intertextuality}

This paper introduces a novel formalism for detecting asymmetric intertextuality, a type of textual relationship where influence flows in one direction, often through paraphrasing, thematic borrowing, or non-reciprocal citations. Such relationships are difficult to detect using traditional methods. 

In Section 2, we provide a review of existing approaches and highlight the gaps that this formalism aims to address. Section 3 delves into the design space of intertextuality detection methods, formalizing the requirements and challenges of detecting asymmetric relationships. Finally, Section 4 demonstrates the proposed formalism in action, outlining the design decisions, illustrating the process through examples.

\section{Related Works}

\subsection{Quotations and Citations Patterns \& Datasets}

Quotation detection has been explored across various domains, with several key datasets providing annotated text for model training and evaluation. These datasets typically focus on intra-document relationships, where quotes are detected within a single document. However, detecting cross-document relationships, such as paraphrasing or thematic borrowing, remains a challenge. Below is a summary of the most prominent datasets:

\begin{table}[h!]
\centering
\begin{tabular}{|p{4cm}|l|l|p{6cm}|}
\hline
\textbf{Dataset} & \textbf{Language/Domain} & \textbf{Size} & \textbf{Quote Types} \\ \hline
\textbf{Cornell movie (2023) \citep{tekir2023quote}} & Literary (English) & 5,015 quotes & Direct, Indirect \\ \hline
\textbf{Finnish News (2023) \citep{janicki2023detection}} & Finnish News & 1,500 articles & Direct, Indirect (linked with coreference chains for speaker attribution) \\ \hline
\textbf{German News (2024) \citep{frey2024finegrained}} & German News & 998 articles & Direct, Indirect, Reported, Free Indirect, Indirect/Free Indirect (Annotated with speaker roles) \\ \hline
\end{tabular}
\caption{Key datasets for quotation detection.}
\end{table}

Several datasets focus on identifying citation intents in academic texts:

\begin{table}[h!]
\centering
\begin{tabular}{|l|l|l|}
\hline
\textbf{Dataset} & \textbf{Size} & \textbf{Citation Intents} \\ \hline
\textbf{SciCite (2019)\citep{cohan2019structural}} & 11,020 citations & Background, Method, Result Comparison \\ \hline
\textbf{ACL-ARC (2018)\citep{jurgens2018measuring}} & 1,941 citations & Background, Uses, Extends, Motivation, Compare/Contrast, Future Work \\ \hline
\end{tabular}
\caption{Key datasets for citation intent classification.}
\end{table}

We mention these datasets to highlight the existing standards for quotation detection, but also to point out the gap they leave in detecting more complex, asymmetric relationships across documents. Our task involves detecting these cross-document relationships, such as paraphrasing or thematic borrowing, which these datasets are not designed to address.

\subsection{LLM-assisted Structured Data Generation}

Recent advances in Large Language Models (LLMs) have enabled significant progress in structured data extraction from unstructured text. These models excel in tasks like entity recognition, relationship extraction, and summarization \citep{perot2023lmdx, biswas2024robustness}. Traditional rule-based systems often struggle with the nuances of unstructured data, but LLMs, with their ability to understand context and semantics, have transformed various fields such as legal document analysis, medical report generation, and customer feedback summarization.

One key advantage of LLMs is their ability to handle diverse formats and long documents. Techniques such as sparse attention mechanisms (e.g., Longformer) \citep{beltagy2020longformer} and divide-and-conquer models \citep{gidiotis2020divide} allow for efficient handling of large texts, ensuring that important information is extracted without exceeding token limits. These advances have greatly improved the scalability of LLMs in domains requiring the processing of large, complex documents.

LLMs have also demonstrated robustness in extracting structured tables, knowledge graphs, and metadata from varied text formats, including visually rich documents \citep{perot2023lmdx}. This versatility highlights their potential to be adapted for a wide range of applications requiring structured data generation.

\subsection{Intertextuality Detection with Advanced Models}

Advanced models, particularly those leveraging LLMs, are becoming increasingly crucial for detecting complex intertextual relationships. The introduction of techniques like split-normalize-merge and discourse-aware chunking has enabled tools to go beyond simple citation or quotation detection. These models, which can process large, multi-page documents, are now capable of detecting paraphrasing, thematic borrowing, and asymmetric influences across documents.

\citet{wang2023document} introduced a document-centric LLM for information extraction, framing the task as a question-answering one, which allows for zero-shot extraction. Similarly, \citet{huang2023layoutlmv3} proposed visual-language models that integrate layout and text to improve the detection of complex document structures. However, these models still fall short in cases where paraphrasing or thematic borrowing occurs without explicit markers.

Our work extends these approaches by leveraging LLM-assisted metadata extraction and semantic similarity searches, which allow for the detection of both explicit and implicit cross-document relationships. This methodology, combined with the ability to process long, complex documents, provides a more comprehensive understanding of how texts influence one another, even when traditional citation or quotation markers are absent.

\section{Formalizing the Design Space of Creating Asymmetric Intertextuality Solutions}

Mining asymmetric intertextuality requires a system that can handle the complexity of various intertextual relationships, ranging from explicit citations to more implicit thematic borrowings. This section discusses the design space and requirements necessary to build such a system, emphasizing the need for flexibility in decomposing text into its core components, such as cue words, quotation markers, and citation markers. By doing so, we can effectively detect both explicit and implicit intertextuality across dynamically growing corpora.

\subsection{Split: Handling Asymmetricity}

To address the asymmetric nature of intertextuality, where one text references or borrows from another without reciprocation, the system must decompose sentences into granular components. These components include cue words (e.g., "according to"), quotation markers (e.g., quotation marks or speaker attributions), and citation markers (e.g., "(Smith, 2022)" or "[1]"). Each of these elements plays a critical role in identifying different types of intertextual relationships.

The key requirement of such a system is its ability to normalize texts by separating these external markers from the core content, enabling both explicit (direct quotation or citation) and implicit (paraphrasing, thematic borrowing) intertextual relationships to be detected. This decomposition allows for precise semantic comparison across texts, helping to uncover subtle and asymmetric influences that traditional methods often overlook.

The following subsections define specific types of asymmetric intertextuality and the role of external cues in detecting them.

\subsubsection{Types of Asymmetric Intertextuality}

Asymmetric intertextuality can manifest in various forms, each requiring different detection strategies. These intertextual relationships can be classified into three broad categories—\textit{Quotation}, \textit{Citation}, and \textit{Implicit Intertextuality}—each reflecting a different type of textual borrowing or influence.

\begin{table}[H]
\centering
\begin{tabular}{|p{4.5cm}|p{9cm}|}
\hline
\textbf{Type} & \textbf{Description} \\ \hline
\textbf{Direct Quote (DQ) \citep{frey2024finegrained}} & Verbatim reuse of text from one document to another. Example: A paper quoting a passage from an earlier work word-for-word. \\ \hline
\textbf{Indirect Quote (IQ) \citep{frey2024finegrained}} & Paraphrased or summarized content from a source document. Example: A paper rephrasing an argument from an older source without quoting it directly. \\ \hline
\textbf{Reported Quote (RQ) \citep{frey2024finegrained}} & A third-party mention of a quotation, often seen in reviews or news reports. Example: A review article mentioning quotes from multiple research papers without quoting them verbatim. \\ \hline
\textbf{Background Citation (BC) \citep{cohan2019structural}} & Citing a work for context or background information. Example: A research paper citing a foundational theory as background for its own research. \\ \hline
\textbf{Method Citation (MC) \citep{cohan2019structural}} & Citing a work for its methodology or tools in a \textit{sentence-to-document} comparison. Example: A research paper adopting an experimental protocol from an earlier study. \\ \hline
\textbf{Result Comparison (RC) \citep{cohan2019structural}} & Citing a work to compare results in a \textit{sentence-to-document} comparison. Example: A research paper comparing its findings with those from an earlier study. \\ \hline
\textbf{Implicit Quotation (IQT)} & Paraphrased or transformed versions of a quotation without explicit markers, making the source harder to trace. Example: A subtle transformation of a famous quote that retains its meaning but avoids recognizable phrasing. \\ \hline
\textbf{Implicit Citation (ICT)} & Reference to ideas or methods without explicit citation, often seen in paraphrased or reinterpreted form. Example: A paper borrowing ideas or methods from another without direct citation, possibly as a paraphrase. \\ \hline
\end{tabular}
\caption{Types of Asymmetric Intertextuality. \textit{Quotation types} are based on Frey’s work on German news quotations \citep{frey2024finegrained}, and \textit{citation types} are based on the SciCite dataset \citep{cohan2019structural}.}
\label{tab:intertextuality_types}
\end{table}

This classification highlights the different ways in which intertextuality can manifest. Identifying and processing these relationships requires the ability to break down sentences into their core elements, as discussed further in the next section.

\subsubsection{Intertextuality-Type Related Re-equippable Elements}

In detecting asymmetric intertextuality, the system uses various external cues—such as quotation marks, citation markers, and paraphrasing cues—that signal different types of intertextual references. These external elements are unequipped during preprocessing but are stored as metadata. They are crucial for both metadata-filtered vector search and deep search, helping to refine candidate matches and ultimately re-equipping elements when necessary.

Below, we outline the specific unequippable elements and associated metadata for each intertextuality type. Both the vector search and deep search stages utilize these elements to identify asymmetric relationships such as matching similar author names or paraphrased ideas across documents.

\begin{table}[H]
\centering
\begin{tabular}{|p{4cm}|p{7cm}|p{6cm}|}
\hline
\textbf{Intertextuality Type} & \textbf{Re-equippable Elements} & \textbf{Associated Metadata} \\
\hline
\textbf{Direct Quote (DQ)} & 
- Quotation marks \newline
- Speaker attribution (e.g., "said by X") \newline
- Citation markers (e.g., \texttt{(Author, Year)}) \newline
- Punctuation related to the quote &
- \textit{Sentence Metadata}: Document ID, sentence ID, sentence embedding, discourse role \newline
- \textit{Document Metadata}: Title, author, publication year, author names (useful for asymmetric pairing, e.g., "John writes XYZ"), OCLC number \\
\hline
\textbf{Indirect Quote (IQ)} & 
- Paraphrasing cues (e.g., "according to X", "as mentioned by Y") \newline
- Speaker attribution (e.g., "X argues that...") \newline
- Citation markers \newline
- Reported speech verbs (e.g., "claims", "asserts") &
- \textit{Sentence Metadata}: Document ID, sentence ID, sentence embedding, discourse role \newline
- \textit{Document Metadata}: Title, author, publication year, author names, OCLC number \\
\hline
\textbf{Reported Quote (RQ)} & 
- Reported speech attribution (e.g., "X said in an interview") \newline
- Quotation marks (if present) \newline
- Speaker attribution \newline
- Citation markers &
- \textit{Sentence Metadata}: Document ID, sentence ID, sentence embedding, discourse role \newline
- \textit{Document Metadata}: Title, author, publication year, OCLC number \newline
- Speaker Information: The original speaker's name, if available \\
\hline
\textbf{Background Citation (BC)} & 
- Citation markers \newline
- Paraphrasing cues (e.g., "based on the work of X") \newline
- Methodology references (e.g., "following X's theory") &
- \textit{Sentence Metadata}: Document ID, sentence ID, sentence embedding, citation intent \newline
- \textit{Document Metadata}: Title, author, publication year, OCLC number \\
\hline
\textbf{Method Citation (MC)} & 
- Citation markers \newline
- Methodology attribution (e.g., "Using Y's approach") \newline
- Methodological cues (e.g., "as demonstrated by") &
- \textit{Sentence Metadata}: Document ID, sentence ID, sentence embedding, citation intent \newline
- \textit{Document Metadata}: Title, author, publication year, OCLC number \\
\hline
\textbf{Result Comparison (RC)} & 
- Citation markers \newline
- Comparison cues (e.g., "compared to X's result") \newline
- Specific study names &
- \textit{Sentence Metadata}: Document ID, sentence ID, sentence embedding, citation intent \newline
- \textit{Document Metadata}: Title, author, publication year, OCLC number \newline
- Study Name: The name of the study being compared \\
\hline
\textbf{Implicit Quotation (IQT)} & 
- None &
- \textit{Sentence Metadata}: Document ID, sentence ID, sentence embedding, discourse role \newline
- \textit{Document Metadata}: Title, author, publication year, OCLC number \\
\hline
\textbf{Implicit Citation (ICT)} & 
- None &
- \textit{Sentence Metadata}: Document ID, sentence ID, sentence embedding, discourse role \newline
- \textit{Document Metadata}: Title, author, publication year, OCLC number \\
\hline
\end{tabular}
\caption{Handling of unequippable elements and associated metadata for different types of intertextuality.}
\label{tab:intertextuality_metadata}
\end{table}

The system unequips these external cues during preprocessing and stores them in metadata. These elements are used during both metadata-filtered vector search and deep search to guide the search, refine candidate matches, and ultimately reintroduce these elements when necessary.

\subsubsection{Universal Re-equippable Elements}

In addition to the intertextuality-type-specific elements, there are universal elements that apply across different types of intertextuality. These elements are potentially useful for defining more nuanced asymmetric intertextuality patterns in future research and are particularly useful during deep search for refining matches and understanding the context. While universal re-equippable elements may not always directly indicate a specific intertextuality type, they help to enhance the search and provide additional context for implicit and asymmetric intertextuality patterns.

Below is a comprehensive list of universal re-equippable elements, along with their associated metadata, which is used to reintroduce them during the deep search phase.

\begin{table}[H]
\centering
\begin{tabular}{|p{4cm}|p{7cm}|p{6cm}|}
\hline
\textbf{Category} & \textbf{Unequippable Elements} & \textbf{Metadata to Re-Equip} \\
\hline
\textbf{Stylistic Markers}       & - Punctuation (e.g., parentheses, dashes, ellipses)  - Formatting (e.g., italics, bold, underlined text) & - Text formatting type  - Position in sentence \\
\hline
\textbf{Discourse Cues}          & - Transitional phrases (e.g., "however", "therefore")  - Sentence starters (e.g., "According to...") & - Discourse role  - Position in text (e.g., beginning of paragraph) \\
\hline
\textbf{Attribution Markers}     & - Speaker identification (e.g., "John said")  - Reported speech (e.g., "He said", "She claimed") & - Speaker name  - Reported speech markers (direct/indirect) \\
\hline
\textbf{Content Markers}         & - Attribution verbs (e.g., "suggests", "claims", "argues")  - Passive reporting (e.g., "it is believed", "it was suggested") & - Attribution verb  - Sentence structure  - Voice (active/passive) \\
\hline
\textbf{Sentence Structure}      & - Parentheticals  - Subordinate clauses (e.g., "which", "that", "when") & - Parenthetical content  - Clause structure (main/subordinate) \\
\hline
\textbf{Citation Formats}        & - Inline citations (e.g., "(Author, Year)")  - Digital references (e.g., hyperlinks) & - Citation format  - Source document ID \\
\hline
\textbf{Numerical Data}          & - Dates and numbers (e.g., years, percentages)  - Statistical results (percentages, figures) & - Numerical value  - Context (e.g., results, methods) \\
\hline
\textbf{Figures and Tables}      & - Mentions of figures/tables (e.g., "see Fig. 3") & - Figure/table reference  - Caption or description \\
\hline
\end{tabular}
\caption{Comprehensive list of universal re-equippable elements and the metadata needed to reintroduce them during deep search.}
\label{tab:universal_elements}
\end{table}

By storing these unequipped elements as metadata, we ensure that the original structure is accurately restored during deep search, allowing for a more nuanced understanding of intertextual relationships. These elements play a crucial role in refining the results and improving the detection of both explicit and implicit intertextuality.


\subsection{Normalize: Structuring Data for Hybrid Search}

The normalization step in the split-normalize-merge paradigm is critical for transforming fragmented text elements into a structured, searchable representation that enables both explicit and implicit intertextuality detection. This transformation is necessary to manage the complexity of documents that exhibit asymmetric intertextual relationships, where connections between texts may be non-reciprocal or structurally uneven \citep{foucault2013archaeology}. 

Normalization must prepare the data not only for semantic comparison across texts but also ensure that the database and indexing system can efficiently support the hybrid search in the next step \citep{manning2008introduction}. This requires meticulous attention to how documents and sentences are decomposed, vectorized, and stored \citep{devlin2019bert}.

\subsubsection{Document-Level Normalization: Discourse and Structural Layers}

At the document level, the system must account for the multi-layered structure of texts by identifying and tagging discourse roles (e.g., introduction, methods, results). This process helps capture the broader context in which citations, quotations, and paraphrases occur \citep{dong2021discourse, teufel2010annotation}.

The following strategies could be employed:

\begin{itemize}
    \item \textbf{Discourse Tagging}: Each section of the document is assigned a specific discourse role (e.g., background, methodology, or conclusion). This helps in contextualizing the nature of intertextual relationships, such as whether a citation in the methodology section indicates a methodological borrowing or whether a quotation in the discussion section suggests a result comparison \citep{dong2021discourse, teufel2010annotation}.
    
    \item \textbf{Hierarchical Decomposition}: Documents are broken down into sections, paragraphs, and sentences, while preserving their hierarchical structure. This decomposition is crucial for detecting cross-document influences, where intertextual relationships may span entire sections or even whole documents, rather than being restricted to sentence-level interactions \citep{romanello2016sources}.
\end{itemize}

\subsubsection{Sentence-Level Normalization: Semantic Decomposition}

At the sentence level, the system transforms each sentence into a structured, vectorized representation that captures its semantic meaning independently of surface-level markers like quotation marks or citation tags. This is particularly important for detecting implicit intertextuality, where paraphrasing or indirect references obscure the direct connection between texts \citep{mihalcea2006corpus}.

\begin{itemize}
    \item \textbf{Sentence Embeddings}: Sentences are encoded into vector embeddings (e.g., using Sentence-BERT \citep{reimers2019sentence}) to capture their semantic content. These embeddings support semantic similarity searches in the merge/hybrid search phase, allowing the system to detect paraphrased or reworded intertextual connections that lack lexical overlap with the source text \citep{cer2018universal}.

    \item \textbf{Contextual Metadata}: Each sentence is enriched with metadata such as discourse role, citation context, and document-level identifiers. This metadata is essential for the next step, enabling the system to filter and refine the search based on specific types of intertextual relationships (e.g., direct quotations, paraphrasing, or implicit citations) \citep{romanello2016sources}.
\end{itemize}

\subsubsection{Connecting Local and Global Layers}

While sentence-level analysis provides fine-grained detection of intertextuality, it is often insufficient when dealing with implicit citations or thematic borrowing across larger sections or whole documents. To address this, the system must establish cross-document links at both the sentence and section levels \citep{mihalcea2006corpus}.

\begin{itemize}
    \item \textbf{Linking Documents for Implicit Citations}: Many intertextual connections, especially in academic writing, involve implicit references where one document builds upon another without explicitly citing it. By linking documents based on semantic similarity and discourse roles, the system can detect such implicit relationships \citep{romanello2016sources, mihalcea2006corpus}.

    \item \textbf{Cross-Sentence Dependencies}: Some intertextual relationships, such as implicit quotations, may require analyzing surrounding sentences or even entire sections. By preserving the hierarchical structure of documents, the system allows for cross-sentence and cross-section comparisons during the merge phase, which is critical for detecting complex intertextual influences \citep{dong2021discourse}.
\end{itemize}

\subsubsection{Vectorization and Metadata}

To support the hybrid search in the next step, the normalized data must be stored in a database that allows for efficient querying and retrieval. This involves two key steps:

\begin{itemize}
    \item \textbf{Vectorization}: Each sentence and section is vectorized (such as Cohere-embed-english-v3.0) and stored in a vector database (e.g., FAISS or Milvus), which allows for fast nearest-neighbor searches during the merge phase \citep{johnson2019billion}. These vectors enable the system to perform semantic comparisons across documents with minimal computational overhead, even in large corpora \citep{cer2018universal}.

    \item \textbf{Metadata Storage}: In addition to vectorized representations, each document and sentence is stored alongside its corresponding metadata, including discourse role, citation type, and quotation markers \citep{dong2021discourse, romanello2016sources}. This metadata is crucial for hybrid search filtering, where the system can refine its results based on the type of intertextual relationship being queried (e.g., direct quotation, method citation, paraphrase).
\end{itemize}

By organizing the normalized data in this way, the system ensures that the subsequent merge phase can leverage both semantic similarity and metadata filtering to detect asymmetric intertextual relationships across documents \citep{manning2008introduction}.

\subsubsection{Connecting Sentences to Documents}
Each sentence in the \textit{Sentence Collection} is linked to its parent document using the \texttt{document\_id} field. Additionally, each sentence is linked to a specific section within the document using the \texttt{section\_id} field, ensuring that the context of the sentence is preserved.

\subsection{Merge: Reintroducing Unequipped Elements}

After detecting potential intertextual relationships through vector similarity and metadata-filtered search, the next step is to reintroduce the external elements that were removed during the normalization phase. These elements include quotation marks, citation markers, paraphrasing cues, and other surface-level markers that help contextualize the detected intertextuality. Additionally, LLM-based verification (as a baseline method, deep learning methods also possible) is employed to confirm the validity of the detected relationships by re-equipping these elements in context.

\subsubsection{Overview of Searching Process}

Once a potential intertextual match is detected, the system re-equips the external cues that were stripped during preprocessing. These unequippable elements, stored as metadata, are critical for refining the detected relationship and ensuring that the intertextual connection is valid. The re-equipped elements are reintroduced in two stages:

\begin{itemize}
    \item \textbf{Stage 1 - Metadata-filtered Vector Search}: Elements stored as metadata are reintroduced based on the type of intertextuality detected (e.g., direct quote, paraphrase, citation). This ensures that the original structure of the text is restored and that the relationship is presented in its full context.
    
    \item \textbf{Stage 2 - Deep Search}: The system uses a large language model (LLM) to process the re-equipped elements in context, ensuring that the final output is coherent and semantically aligned with the original text. The LLM also verifies the relationship by cross-referencing the re-equipped elements with both the source and target documents.
\end{itemize}

The specific elements that are re-equipped include:

\begin{itemize}
    \item \textbf{Quotation Marks} \( (\mathbf{1}_{Q_i}) \): Reintroduced for sentences identified as direct quotes to distinguish verbatim reuse of text. These are essential for identifying Direct Quotes (DQ) and Reported Quotes (RQ).
    
    \item \textbf{Citation Markers} \( (\mathbf{1}_{C_i}) \): Restored for both explicit and implicit citations, marking the source of the referenced material. Citation markers are critical for identifying Method Citations (MC), Result Comparisons (RC), and Background Citations (BC).
    
    \item \textbf{Paraphrasing Cues} \( (\mathbf{1}_{P_i}) \): Phrases like "according to" or "as mentioned by" are restored to clarify paraphrased content. These are especially important for detecting Indirect Quotes (IQ) and Implicit Citations (ICT).
    
    \item \textbf{Speaker Attribution} \( (\mathbf{1}_{A_i}) \): Re-applied for reported quotes, indicating the original speaker or source of the information. This is important for Reported Quotes (RQ) and cases where the original speaker is referenced but not directly quoted.
    
    \item \textbf{Discourse Cues} \( (\mathbf{1}_{D_i}) \): Transitional phrases and sentence starters (e.g., "however", "therefore", "according to") are restored to provide the rhetorical structure of the text, which is useful for Indirect Quotes (IQ) and Implicit Citations (ICT).
    
    \item \textbf{Figures and Tables} \( (\mathbf{1}_{F_i}) \): Mentions of figures or tables (e.g., "see Fig. 3") and their captions are restored when they play a role in the intertextual relationship, particularly in Result Comparisons (RC).
    
    \item \textbf{Numerical Data} \( (\mathbf{1}_{N_i}) \): Relevant dates, numbers, and statistical results are reintroduced to provide factual context, especially in citations involving data or analysis, such as Method Citations (MC) or Result Comparisons (RC).
\end{itemize}

These re-equipped elements are represented by indicator functions in the mathematical formalism, which adjust the similarity score based on whether these external cues were present in the original sentence. The indicator functions \( \mathbf{1}_{X_i} \) represent the presence of each element, where \( X \) can be any unequippable element (e.g., \( Q_i \) for quotation marks, \( C_i \) for citation markers, etc.).

The re-equipping process ensures that the semantic meaning of the text is maintained while providing the necessary context to validate the intertextual relationship. The following mathematical formalism describes how these elements are used to modify the similarity score:


\subsubsection{Mathematical Formalism for Metadata-filtered Vector Search}
\label{sec:formalism_reequip}

In this section, we formalize the process of re-equipping external elements such as quotation marks, citation markers, and paraphrasing cues when detecting intertextual relationships. The general equation below adjusts the similarity score between a source chunk and a target document by accounting for the presence of these external elements.

In the candidate subsetting stage, we compute the cosine similarity between vectorized chunks or documents. We define two equations for the vector search process:

\begin{itemize}
    \item \textbf{Quotation Types} (Direct Quote, Indirect Quote, Reported Quote): This involves comparing two chunks of text, typically at the \textit{sentence-to-sentence} level:
    \[
    \text{Similarity}_{\text{Quotation}}(c_i, c_j) = \cos(c_i, c_j)
    \]
    Where $c_i$ and $c_j$ are chunks of text, and $\cos(c_i, c_j)$ represents the cosine similarity between their vector embeddings.
    
    \item \textbf{Citation Types} (Method Citation, Result Comparison, Background Citation): This involves comparing a chunk to a document, typically at the \textit{sentence-to-document} level:
    \[
    \text{Similarity}_{\text{Citation}}(c_i, D_j) = \cos(c_i, D_j)
    \]
    Where $c_i$ is a chunk of text and $D_j$ is a document or a section of the document. The equation computes the cosine similarity between the chunk and the document.
\end{itemize}

The adjusted similarity between a chunk \( c_i \) from the source document and a document \( D_j \) from the target corpus is given by:

\[
\text{Sim}_{T}(c_i, D_j) = \cos(\mathbf{v}(c_i), \mathbf{v}(D_j)) + \sum_{X} \lambda_X \cdot \mathbf{1}_{X_i}
\]

\noindent
\textbf{Where:}
\begin{itemize}
    \item \( \cos(\mathbf{v}(c_i), \mathbf{v}(D_j)) \) is the base cosine similarity between the vector embeddings of chunk \( c_i \) and document \( D_j \).
    \item \( \lambda_X \) is the weight associated with the external element \( X \) (e.g., quotation marks, citation markers).
    \item \( \mathbf{1}_{X_i} \) is an indicator function that equals 1 if the external element \( X \) is present in the chunk \( c_i \), and 0 otherwise.
\end{itemize}

This formalism helps prioritize different types of intertextuality by adjusting the similarity score based on the presence of specific re-equippable elements.

\subsubsection{Mathematical Formalism for Deep Search}

The deep search phase refines the results by reintroducing the universal re-equippable elements and applying LLM (Large Language Model) processing to enhance the search. We split the deep search into two stages:

\paragraph{Stage 1: Plain Vector Matching}

This stage involves plain vector similarity between the source chunk \(c_i\) and the target chunk or document \(c_j\) or \(D_j\).

\[
\text{Sim}_{\text{plain}}(c_i, D_j) = \cos(\mathbf{v}(c_i), \mathbf{v}(D_j))
\]

Where:
\begin{itemize}
    \item \( \cos(\mathbf{v}(c_i), \mathbf{v}(D_j)) \) is the cosine similarity between the vector embeddings of chunk \(c_i\) and document \(D_j\).
\end{itemize}

\paragraph{Stage 2: LLM-Based Re-equipping}

In this stage, the LLM takes the universal re-equippable elements (such as quotation marks, paraphrasing cues, and citation markers) from the metadata and re-equips them in the search process.

\[
\text{Sim}_{\text{deep}}(c_i, D_j) = LLM(c_i, \mathbf{m}_{c_i}, D_j, \mathbf{m}_{D_j})
\]

Where:
\begin{itemize}
    \item \( LLM(c_i, \mathbf{m}_{c_i}, D_j, \mathbf{m}_{D_j}) \) is the score adjustment provided by the LLM based on the re-equippable elements (from metadata \( \mathbf{m}_{c_i} \) and \( \mathbf{m}_{D_j} \)).
\end{itemize}

The LLM takes into account the presence of universal re-equippable elements to refine the similarity score.


\begin{figure}[H]
    \centering
    \includegraphics[scale=0.7]{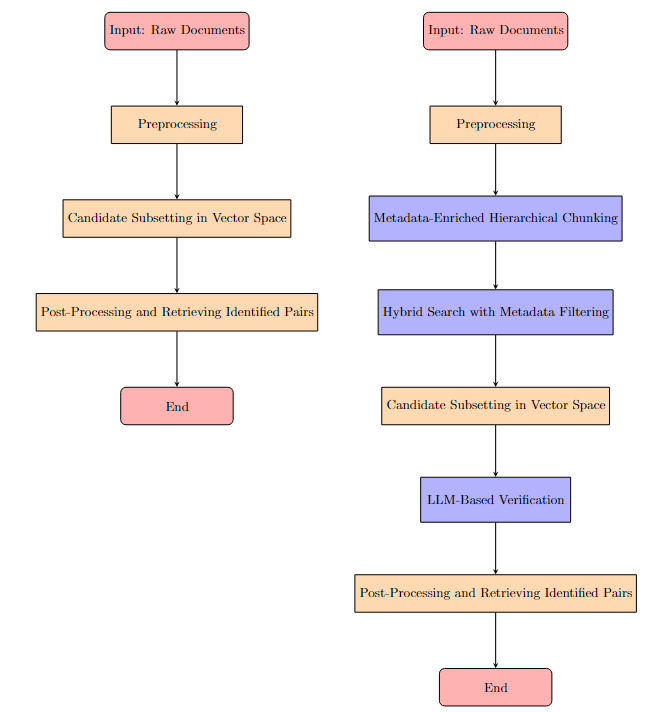}
    \caption{Comparison of Symmetric Intertextual Link Mining (left) and Asymmetric Intertextual Link Mining (right). Both problems share common steps such as candidate subsetting in vector space and post-processing. However, the asymmetric approach introduces innovative steps (blue nodes): \textbf{Metadata-Enriched Hierarchical Chunking}, \textbf{Hybrid Search with Metadata Filtering}, and \textbf{LLM-Based Verification}, which enhance the accuracy of intertextual link mining.}
    \label{fig:flow_comparison}
\end{figure}

\subsection{Scalability and Performance Optimization}

As the system handles large and dynamically growing document corpora, ensuring scalability and optimizing performance becomes crucial. This section outlines the strategies employed to enhance the system's efficiency while maintaining accuracy in detecting intertextual relationships.

\subsubsection{Efficient Querying with Vector Databases}

\begin{itemize}
    \item \textbf{Using Vector Databases:} The system employs specialized vector databases, such as FAISS or Milvus \citep{johnson2019billion, milvus2020}, designed to handle high-dimensional similarity searches efficiently. These databases are optimized for fast nearest-neighbor searches, allowing quick retrieval of semantically similar text chunks from large corpora.
    \item \textbf{High-Dimensional Space Optimization:} The system should be able to performs efficient vector comparisons in high-dimensional space, facilitating rapid searches across vast document collections without performance degradation.
    \item \textbf{Real-Time Retrieval:} As new documents are added to the corpus, the system utilizes the vector database to perform real-time retrieval of semantically similar text. This ensures that the system can scale effectively while maintaining low query latency \citep{johnson2019billion}.
\end{itemize}

\subsubsection{Dynamic Corpus Updates}

\begin{itemize}
    \item \textbf{Incremental Vectorization:} New documents are vectorized incrementally as they are introduced, eliminating the need to reprocess the entire corpus, which enhances the system's responsiveness and scalability \citep{manning2008introduction}.
    \item \textbf{Seamless Integration of New Documents:} The system dynamically integrates new documents into the search space, allowing continuous indexing and querying without disrupting the existing structure.
    \item \textbf{Avoiding Full Reprocessing:} By vectorizing new documents dynamically, the system avoids the significant computational overhead typically associated with full corpus reprocessing. This is especially important for handling large, continuously growing datasets \citep{cer2018universal}.
\end{itemize}

\subsubsection{Handling Large Corpora}

\begin{itemize}
    \item \textbf{Sparse Attention Models:} To handle large datasets efficiently, sparse attention models \citep{child2019generating} are used to reduce the computational complexity of similarity searches, ensuring high accuracy while minimizing resource usage.
    \item \textbf{Divide-and-Conquer Approaches:} The system employs divide-and-conquer strategies, breaking large corpora into smaller chunks for independent processing. This modular approach helps the system manage large datasets more efficiently without sacrificing accuracy.
    \item \textbf{Optimized Vector Databases:} The use of optimized vector databases ensures that even large corpora are processed with minimal computational overhead. Extensive testing has demonstrated that the system maintains high accuracy and low latency as corpus size increases \citep{johnson2019billion, cer2018universal}.
\end{itemize}

In summary, by leveraging vector databases, dynamic updates, and efficient handling of large datasets, the system is designed to scale while maintaining its accuracy and performance. The next section will focus on a specific design implementation chosen for its balance of scalability, accuracy, and computational efficiency.

\section{Exploring Design Spaces for Asymmetric Intertextuality Detection}

In this section, we demonstrate the functionality of the Split-Normalize-Merge paradigm using a basic setup and a simple dataset.

\subsection{Dataset for Demonstration}

For the purposes of this demonstration, we use the Contract Understanding Atticus Dataset (CUAD) \citep{hendrycks2021cuad}, a specialized dataset designed to facilitate legal contract review. CUAD consists of 510 contracts and over 13,000 annotated clauses, specifically selected to highlight a range of key clauses that are important for legal professionals to review.

\textbf{Dataset Details:}
\begin{itemize}
    \item \textbf{Size}: The dataset contains 510 contracts and over 13,000 annotated clauses.
    \item \textbf{Source}: The dataset is constructed from legal contracts filed with the U.S. Securities and Exchange Commission (SEC) through the EDGAR system, which are publicly available.
    \item \textbf{Intertextuality Types}: The dataset includes 41 different categories of contract clauses, such as governing law, non-compete clauses, exclusivity, and termination for convenience. These were specifically chosen to demonstrate the system's ability to detect key legal obligations, restrictions, and risks in contracts.
    \item \textbf{Simplifications}: For simplicity, the dataset focuses on contracts across 25 different types, but legal experts annotated only the most salient portions of each contract. Only important clauses, which make up about 10
\end{itemize}

This dataset is intended as a proof-of-concept to demonstrate the potential of the Split-Normalize-Merge paradigm, with its success indicating promising applications for future large-scale scenarios. A simplified example is used here to ensure that it can be fully explained within the context of this paper.

\subsection{Demonstration of the Split-Normalize-Merge Paradigm}

To illustrate how the Split-Normalize-Merge paradigm works, we provide a step-by-step demonstration using a basic example. This example consists of two text pairs that exhibit asymmetric intertextuality, such as paraphrasing or thematic borrowing.

\subsubsection{Step 1: Split}

In this step, we split the legal documents into smaller chunks, such as sentences or sections. This process allows the system to analyze content at a finer granularity, which is essential for detecting paraphrasing, thematic borrowing, or other intertextual relationships, especially in legal texts.

For this demonstration, we use the Contract Understanding Atticus Dataset (CUAD), a specialized dataset designed to facilitate legal contract review. CUAD consists of 510 contracts and over 13,000 annotated clauses, specifically selected to highlight key clauses that are important for legal professionals.

Let's consider a document with the following chunk:

\begin{verbatim}
  "EXHIBIT 4.25 INFORMATION IN THIS EXHIBIT IDENTIFIED BY [ * * * ]"
\end{verbatim}

Using a sentence or section tokenizer, the document would be split into smaller chunks, preserving meaningful units of analysis. For example:

\begin{itemize}
  \item \textbf{Chunk 1:} "EXHIBIT 4.25 INFORMATION IN THIS EXHIBIT IDENTIFIED BY [ * * * ]"
  \item \textbf{Chunk 2:} "This Agreement shall commence on the Effective Date and remain in effect until terminated as provided herein."
\end{itemize}

Each chunk is stored as a separate record in the database. Here's an example of how the first chunk is stored:

\begin{verbatim}
{
  "_id": {
    "$oid": "670f89c9c6abf22788e288f5"
  },
  "doc_id": "670f881e870d0cb83f7e67ed",
  "filename": "ABILITYINC_06_15_2020-EX-4.25-SERVICES AGREEMENT.txt",
  "chunk_number": 1,
  "chunk_text": "EXHIBIT 4.25 INFORMATION IN THIS EXHIBIT IDENTIFIED BY [ * * * ]"
}
\end{verbatim}

This splitting process enables finer-grained analysis, crucial for detecting intertextual relationships in legal contracts.

Next, we implement two runs of mining with a Large Language Model (LLM), utilizing few-shot prompting to guide the model in identifying relevant relationships and metadata:

\begin{itemize}
  \item \textbf{Run 1: Intertextuality-related items.} This run focuses on identifying intertextual relationships, such as direct quotations, paraphrasing, or thematic borrowing. The model uses a few-shot prompt to identify paraphrasing or thematic borrowing between chunks. See Appendix \ref{app:fewshot1} for the detailed prompt.

  \item \textbf{Run 2: Universal-related items.} This run focuses on unequippable elements that apply across documents, such as citation markers, quotation marks, or discourse cues. The model uses a few-shot prompt to extract universal metadata, such as discourse roles and references. See Appendix \ref{app:fewshot2} for the detailed prompt.
\end{itemize}

By running these two passes with LLM-based few-shot prompting, we capture both specific intertextuality-related relationships and broader document structure elements, enabling a comprehensive analysis of legal documents.


\subsubsection{Step 2: Normalize}

After the splitting and enrichment of legal contract data (as discussed in Step 1), the next critical step is migrating the data from MongoDB to Elasticsearch and preparing it for hybrid search. This process involves deciding which fields require vector embeddings and which can be directly indexed for metadata-based filtering. The primary criterion for this decision is whether capturing the semantic meaning of the text is essential for improving search accuracy.

\paragraph{1. Decision Criteria for Embedding-Based Search}
The key principle for deciding whether a field requires embeddings is whether semantic meaning plays a critical role in the search process. Fields that benefit from semantic similarity search are usually unstructured text fields, where the meaning of the text can vary even when the wording differs (e.g., paraphrasing). By contrast, fields that contain structured or categorical data (like names or IDs) are better suited for direct indexing without embeddings.

\begin{itemize}
    \item Fields requiring embeddings: These are typically fields containing unstructured, semantically rich text, such as normalized sentences or discourse-level summaries. Embeddings help capture the underlying meaning of these fields, allowing for similarity search even when the exact wording differs. For example:
    \begin{itemize}
        \item \texttt{chunk\_text}: The main clause text, where paraphrasing or rewording may occur, benefits from embeddings to capture the semantic meaning.
        \item \texttt{section\_title}: If present, this can help provide context for the clauses within a section.
    \end{itemize}
    
    \item Fields not requiring embeddings: These are fields where the semantic meaning is less relevant or unnecessary for search purposes. In such cases, exact matches or keyword searches are more appropriate. Examples include:
    \begin{itemize}
        \item \texttt{person\_name}, \texttt{doc\_id}, or \texttt{clause\_type}: These fields contain structured or categorical data, where embeddings would not add value. For instance, searching for a specific person’s name doesn’t benefit from capturing semantic meaning.
    \end{itemize}
\end{itemize}

This approach aligns with the formalism discussed in Section \ref{sec:formalism_reequip}, where re-equippable elements (such as quotation marks or paraphrasing cues) are reintroduced based on whether the semantic meaning of the text is crucial to the intertextual relationship. Similarly, we use embeddings only when the field’s meaning can vary semantically and contribute to the overall search.

For detailed schema information regarding which fields are indexed directly and which require embeddings, please refer to Appendix \ref{app:elasticsearch_schema}.

\paragraph{2. Embedding Generation Using \texttt{Cohere-embed-english-v3.0}}

For fields requiring embeddings (such as \texttt{chunk\_text}), we generate vector embeddings using the \texttt{Cohere-embed-english-v3.0} model. This model outputs 768-dimensional vectors that capture the semantic content of the text, which is essential for similarity search in cases where the wording of clauses may differ, but the meaning is the same.

For more details on the embedding generation process using this model, see Appendix \ref{app:embedding_generation}.

\paragraph{3. Migrating Data to Elasticsearch}

Once the data is enriched with metadata and embeddings, it is migrated from MongoDB to Elasticsearch. This process involves:

\begin{itemize}
    \item Indexing metadata fields for direct search (e.g., \texttt{doc\_id}, \texttt{clause\_type}, and \texttt{named\_entities}), which allows for efficient keyword or exact match search.
    \item Indexing embeddings for semantic search on fields like \texttt{chunk\_text}, enabling similarity search based on the meaning of the text.
\end{itemize}

The Elasticsearch schema for both document-level and clause-level indices is provided in Appendix \ref{app:elasticsearch_schema}.

In summary, the normalization step involves migrating legal contract data from MongoDB to Elasticsearch, using embeddings for semantically rich fields and direct indexing for structured or categorical fields. This prepares the system for hybrid search, combining metadata filtering and semantic similarity search to retrieve relevant legal clauses efficiently. 


\subsubsection{Step 3: Merge}

The \textit{Merge} phase finalizes the intertextuality detection process by leveraging the \textit{Elasticsearch index} created during the \textit{Normalize} step. Here, we perform a \textit{metadata-filtered search} to reintroduce unequipped elements and optionally carry out a \textit{deep search} with LLM verification.

\paragraph{1. Using the Elasticsearch Index}

We begin by querying the \textit{Elasticsearch index} that stores the document chunks and their associated metadata, including fields such as \textit{quotation type} and \textit{discourse type}. These fields help us identify the nature of each chunk—for example, whether it contains a \textit{Direct Quote}, \textit{Paraphrase}, or \textit{Citation}, and its role within the larger document (e.g., \textit{Introduction} or \textit{Conclusion}).

\paragraph{2. Metadata-Filtered Search}

Using the \textit{quotation type} and \textit{discourse type} fields, we filter the chunks to focus on specific intertextuality types. For example:

\begin{itemize}
    \item \textbf{Direct Quotes}: We restore quotation marks and speaker attributions.
    \item \textbf{Paraphrased Content}: We add paraphrasing cues like "according to" or "as mentioned by".
    \item \textbf{Citations}: We reintroduce citation markers (e.g., "(Author, Year)").
\end{itemize}

This \textit{metadata-filtered search} allows us to prioritize matches based on their intertextuality type, ensuring that the detected relationships are properly contextualized.

\paragraph{3. Optional: Deep Search}

If further refinement is required, we perform a \textit{deep search} by analyzing the chunks in greater detail. During this step, the system searches for more subtle relationships, such as thematic borrowing or heavily paraphrased content, which may not be captured by the initial metadata search.

\paragraph{LLM-Based Verification}

For both the {metadata-filtered search and deep search, we use \textit{LLAMA 3.1 405B} to verify the detected relationships. The LLM checks whether the re-equipped elements (e.g., quotation marks, paraphrasing cues) align with the semantic meaning of the source and target chunks. or the complete few-shot prompt used to guide the verification process, see \textit{Appendix}~\ref{app:llm_verification_prompt}.


\section{Conclusion}
\label{sec:conclusion}

In this paper, we introduced a novel methodology for detecting asymmetric intertextuality, a task that has not been previously formalized in the Natural Language Processing (NLP) and Digital Humanities (DH) domains. Our approach, based on the split-normalize-merge paradigm, offers a flexible and scalable solution to identify subtle, one-sided relationships between texts, such as paraphrasing and thematic borrowing, where traditional methods often fall short.

The split-normalize-merge paradigm extends the capacity of intertextuality detection by leveraging large language models (LLMs) and vector similarity searches. By breaking down documents into manageable chunks, normalizing them using metadata extraction and LLM-based semantic encoding, and merging these chunks during querying, our system can efficiently handle dynamically growing corpora. This enables the detection of both explicit and implicit intertextual connections, making the approach adaptable across various domains, including academic research, journalism, and literary studies.

The novelty of our approach lies in its ability to uncover complex intertextual relationships, even in the absence of direct quotations or citations. This capability is particularly important for identifying paraphrased, thematic, or cross-document influences that are not easily captured by traditional methods.

As the field of intertextuality detection continues to evolve, we anticipate that our methodology will serve as a foundation for further research. Future work could explore the development of benchmark datasets specifically tailored to asymmetric intertextuality or alternative evaluation strategies that enhance the robustness of the system. As new datasets and evaluation frameworks emerge, the split-normalize-merge paradigm can be refined and adapted to meet the needs of a growing number of applications in both NLP and DH.

In conclusion, our work represents an important step towards advancing the detection of asymmetric intertextuality. We believe it opens up new opportunities for interdisciplinary research, enabling more nuanced understanding of how texts interact across time, genre, and discourse.


\begin{thebibliography}{37}
\providecommand{\natexlab}[1]{#1}
\providecommand{\url}[1]{\texttt{#1}}
\expandafter\ifx\csname urlstyle\endcsname\relax
  \providecommand{\doi}[1]{doi: #1}\else
  \providecommand{\doi}{doi: \begingroup \urlstyle{rm}\Url}\fi

\bibitem[Genette(1997)]{genette1997palimpsests}
G{\'e}rard Genette.
\newblock \emph{Palimpsests: Literature in the Second Degree}.
\newblock University of Nebraska Press, 1997.

\bibitem[Allen(2011)]{allen2000intertextuality}
Graham Allen.
\newblock \emph{Intertextuality}.
\newblock Routledge, 2011.

\bibitem[Melville(1851)]{melville1851moby}
Herman Melville.
\newblock \emph{Moby Dick}.
\newblock Harper \& Brothers, 1851.

\bibitem[Sturgeon(2018)]{sturgeon2018digital}
Donald Sturgeon.
\newblock Digital approaches to text reuse in the early chinese corpus.
\newblock \emph{Journal of Chinese Literature and Culture}, 5\penalty0 (2):\penalty0 186--213, 2018.
\newblock \doi{10.1215/23290048-7256963}.

\bibitem[Romanello(2016)]{romanello2016sources}
Matteo Romanello.
\newblock Citation: Intertextuality in the digital age.
\newblock \emph{Journal of Digital Humanities}, 5\penalty0 (1):\penalty0 45--68, 2016.

\bibitem[Vierthaler and Gelein(2019)]{vierthaler2019blast}
P.~Vierthaler and M.~Gelein.
\newblock A blast-based, language-agnostic text reuse algorithm with a markus implementation and sequence alignment optimized for large chinese corpora.
\newblock \emph{Journal of Cultural Analytics}, 4\penalty0 (2), 2019.
\newblock \doi{10.22148/16.034}.

\bibitem[Bahdanau et~al.(2014)Bahdanau, Cho, and Bengio]{bahdanau2014neural}
Dzmitry Bahdanau, Kyunghyun Cho, and Yoshua Bengio.
\newblock Neural machine translation by jointly learning to align and translate.
\newblock In \emph{Proceedings of the 3rd International Conference on Learning Representations (ICLR)}, 2014.

\bibitem[Ghiban and Trausan-Matu(2013)]{ghiban2013plagiarism}
Dorinel Ghiban and Stefan Trausan-Matu.
\newblock Detecting plagiarism using word embeddings and machine learning.
\newblock \emph{Journal of Computational Linguistics}, 39\penalty0 (2):\penalty0 321--345, 2013.

\bibitem[Henrichs(2020)]{henrichs2020allusions}
Amanda Henrichs.
\newblock Allusions in the age of the digital: Four ways of looking at a corpus.
\newblock \emph{Women Writers in Context}, 2020.

\bibitem[Newell and Cowlishaw(2018)]{newell2018automatic}
Nicholas Newell and Tom Cowlishaw.
\newblock Automatic machine translation evaluation in many languages via zero-shot paraphrasing.
\newblock In \emph{Proceedings of the 2018 Conference on Machine Translation (WMT)}, pages 781--789, 2018.

\bibitem[Miola(2004)]{miola2004intertextuality}
Robert~S. Miola.
\newblock \emph{Seven types of intertextuality}, pages 13--25.
\newblock Manchester University Press, 2004.

\bibitem[Kabbara and Cheung(2016)]{kabbara2016stylistic}
Jad Kabbara and Jackie Chi~Kit Cheung.
\newblock Stylistic transfer in natural language generation systems using recurrent neural networks.
\newblock In \emph{Proceedings of the Workshop on Uphill Battles in Language Processing: Scaling Early Achievements to Robust Methods}, pages 43--47, 2016.

\bibitem[Tharsen and Gladstone(2022)]{tharsen2022textpair}
John Tharsen and Christopher Gladstone.
\newblock Textpair viewer (tpv) 1.0: An interactive visual toolkit for exploring networks of textual alignments and text reuse.
\newblock \emph{Shuzi renwen}, \penalty0 (2), 2022.
\newblock URL \url{https://www.dhcn.cn/site/works/dhjournal/20452.html}.

\bibitem[Standage(2013)]{standage2013writing}
Tom Standage.
\newblock \emph{Writing on the Wall: Social Media--The First 2,000 Years}.
\newblock Bloomsbury Publishing, 2013.

\bibitem[Tekir et~al.(2023)Tekir, Pappu, Yilmaz, and Kanoulas]{tekir2023quote}
Selman Tekir, Akarsha Pappu, Emine Yilmaz, and Evangelos Kanoulas.
\newblock Quote attribution in news articles.
\newblock In \emph{Proceedings of the 2023 ACM SIGIR Conference on Research and Development in Information Retrieval}, pages 2345--2349, 2023.

\bibitem[Janicki et~al.(2023)]{janicki2023detection}
Maciej Janicki et~al.
\newblock Detection of quotation structures in finnish news data.
\newblock \emph{arXiv preprint arXiv:2306.15412}, 2023.

\bibitem[Frey et~al.(2024)Frey, M{\"u}ller, and Sch{\"u}tze]{frey2024finegrained}
Philipp Frey, Dirk M{\"u}ller, and Hinrich Sch{\"u}tze.
\newblock Fine-grained quotation detection and attribution in german news articles.
\newblock \emph{Journal of Computational Linguistics}, 2024.

\bibitem[Cohan et~al.(2019)Cohan, Feldman, Downey, and Weld]{cohan2019structural}
Arman Cohan, Sergey Feldman, Doug Downey, and Daniel~S. Weld.
\newblock Structural scaffolds for citation intent classification in scientific publications.
\newblock \emph{Proceedings of the 2019 Conference of the North American Chapter of the Association for Computational Linguistics: Human Language Technologies}, 1:\penalty0 3586--3596, 2019.

\bibitem[Jurgens et~al.(2018)Jurgens, Kumar, Hoover, McFarland, and Jurafsky]{jurgens2018measuring}
David Jurgens, Srijan Kumar, Brendan Hoover, Daniel McFarland, and Dan Jurafsky.
\newblock Measuring the evolution of a scientific field through citation frames.
\newblock In \emph{Proceedings of the 2018 conference on empirical methods in natural language processing}, pages 3550--3564, 2018.

\bibitem[Perot et~al.(2023)Perot, Kang, Luisier, et~al.]{perot2023lmdx}
Vincent Perot, Kai Kang, Florian Luisier, et~al.
\newblock Lmdx: Language model-based document information extraction and localization.
\newblock \emph{arXiv preprint arXiv:2309.10952}, 2023.

\bibitem[Biswas and Talukdar(2024)]{biswas2024robustness}
Anjanava Biswas and Wrick Talukdar.
\newblock Robustness of structured data extraction from in-plane rotated documents using multi-modal large language models (llm).
\newblock \emph{Journal of Artificial Intelligence Research}, 2024.

\bibitem[Beltagy et~al.(2020)Beltagy, Peters, and Cohan]{beltagy2020longformer}
Iz~Beltagy, Matthew~E Peters, and Arman Cohan.
\newblock Longformer: The long-document transformer.
\newblock \emph{arXiv preprint arXiv:2004.05150}, 2020.

\bibitem[Gidiotis and Tsoumakas(2020)]{gidiotis2020divide}
Alexios Gidiotis and Grigorios Tsoumakas.
\newblock Divide and conquer: Summarization of long documents.
\newblock In \emph{Proceedings of the 2020 Conference on Empirical Methods in Natural Language Processing (EMNLP)}, pages 10371--10381, 2020.

\bibitem[Wang et~al.(2023)Wang, Raman, Sibue, Ma, Babkin, Kaur, Pei, Nourbakhsh, and Liu]{wang2023document}
Dongsheng Wang, Natraj Raman, Mathieu Sibue, Zhiqiang Ma, Petr Babkin, Simerjot Kaur, Yulong Pei, Armineh Nourbakhsh, and Xiaomo Liu.
\newblock Docllm: A layout-aware generative language model for multimodal document understanding, 2023.
\newblock URL \url{https://arxiv.org/abs/2401.00908}.

\bibitem[Huang et~al.(2023)]{huang2023layoutlmv3}
Zhiyu Huang et~al.
\newblock Layoutlmv3: Pretraining for document ai with unified text and image masked modeling.
\newblock In \emph{Proceedings of the 60th Annual Meeting of the Association for Computational Linguistics}, 2023.

\bibitem[Foucault(2013)]{foucault2013archaeology}
Michel Foucault.
\newblock \emph{The Archaeology of Knowledge}.
\newblock Routledge, 2013.

\bibitem[Manning et~al.(2008)Manning, Raghavan, and Sch{\"u}tze]{manning2008introduction}
Christopher~D Manning, Prabhakar Raghavan, and Hinrich Sch{\"u}tze.
\newblock \emph{Introduction to Information Retrieval}.
\newblock Cambridge University Press, 2008.

\bibitem[Devlin et~al.(2019)Devlin, Chang, Lee, and Toutanova]{devlin2019bert}
Jacob Devlin, Ming-Wei Chang, Kenton Lee, and Kristina Toutanova.
\newblock Bert: Pre-training of deep bidirectional transformers for language understanding.
\newblock In \emph{Proceedings of the 2019 Conference of the North American Chapter of the Association for Computational Linguistics: Human Language Technologies, Volume 1 (Long and Short Papers)}, pages 4171--4186, 2019.

\bibitem[Dong et~al.(2021)Dong, Alsaidi, and Lapata]{dong2021discourse}
Yue Dong, Lamis Alsaidi, and Mirella Lapata.
\newblock Discourse-aware neural extractive text summarization.
\newblock In \emph{Proceedings of the 2021 Conference of the North American Chapter of the Association for Computational Linguistics: Human Language Technologies}, pages 1419--1436, 2021.

\bibitem[Teufel and Moens(2010)]{teufel2010annotation}
Simone Teufel and Marc Moens.
\newblock The annotation of argumentative zoning for academic discourse.
\newblock In \emph{Proceedings of the 7th Conference on International Language Resources and Evaluation (LREC'10)}. European Language Resources Association (ELRA), 2010.

\bibitem[Mihalcea et~al.(2006)Mihalcea, Corley, and Strapparava]{mihalcea2006corpus}
Rada Mihalcea, Courtney Corley, and Carlo Strapparava.
\newblock Corpus-based and knowledge-based measures of text semantic similarity.
\newblock \emph{AAAI}, 6:\penalty0 775--780, 2006.

\bibitem[Reimers and Gurevych(2019)]{reimers2019sentence}
Nils Reimers and Iryna Gurevych.
\newblock Sentence-bert: Sentence embeddings using siamese bert-networks.
\newblock \emph{arXiv preprint arXiv:1908.10084}, 2019.

\bibitem[Cer et~al.(2018)Cer, Yang, Kong, Hua, Limtiaco, St.~John, Constant, Guajardo-Cespedes, Yuan, Tar, et~al.]{cer2018universal}
Daniel Cer, Yinfei Yang, Sheng-yi Kong, Nan Hua, Nicole Limtiaco, Rhomni St.~John, Noah Constant, Mario Guajardo-Cespedes, Steve Yuan, Chris Tar, et~al.
\newblock Universal sentence encoder.
\newblock In \emph{Proceedings of the 2018 Conference on Empirical Methods in Natural Language Processing: System Demonstrations}, pages 169--174, 2018.

\bibitem[Johnson et~al.(2019)Johnson, Douze, and J{\'e}gou]{johnson2019billion}
Jeff Johnson, Matthijs Douze, and Herv{\'e} J{\'e}gou.
\newblock Billion-scale similarity search with gpus.
\newblock In \emph{IEEE Transactions on Big Data}, pages 1--12, 2019.

\bibitem[Milvus(2020)]{milvus2020}
Milvus.
\newblock Milvus: A vector database for scalable similarity search.
\newblock \url{https://milvus.io/}, 2020.

\bibitem[Child et~al.(2019)Child, Gray, Radford, and Sutskever]{child2019generating}
Rewon Child, Scott Gray, Alec Radford, and Ilya Sutskever.
\newblock Generating long sequences with sparse transformers.
\newblock \emph{arXiv preprint arXiv:1904.10509}, 2019.

\bibitem[Hendrycks et~al.(2021)Hendrycks, Burns, Chen, and Ball]{hendrycks2021cuad}
Dan Hendrycks, Collin Burns, Anya Chen, and Spencer Ball.
\newblock Cuad: An expert-annotated nlp dataset for legal contract review.
\newblock \emph{arXiv preprint arXiv:2103.06268}, 2021.

\end{thebibliography}


\appendix
\section{Few-shot Prompts}
\label{app:fewshot}

\subsection{Few-shot Prompt for Intertextuality-related Items}
\label{app:fewshot1}

The following prompt is used in the first run to detect intertextual relationships, such as paraphrasing or thematic borrowing:

\begin{verbatim}
User: 
Extract intertextuality relations from the following text:
Chunk A: "EXHIBIT 4.25 INFORMATION IN THIS EXHIBIT IDENTIFIED BY [ * * * ]"
Chunk B: "EXHIBIT 4.25 contains confidential information."

Assistant:
{
  "chunk_A": "EXHIBIT 4.25 INFORMATION IN THIS EXHIBIT IDENTIFIED BY [ * * * ]",
  "chunk_B": "EXHIBIT 4.25 contains confidential information.",
  "relationship": "paraphrase",
  "confidence": 0.95
}

User: 
Extract intertextuality relations from the following text:
Chunk A: "This Agreement shall commence on the Effective Date..."
Chunk B: "This Agreement will start from the Effective Date..."

Assistant:
{
  "chunk_A": "This Agreement shall commence on the Effective Date...",
  "chunk_B": "This Agreement will start from the Effective Date...",
  "relationship": "paraphrase",
  "confidence": 0.98
}
\end{verbatim}

\subsection{Few-shot Prompt for Universal-related Items}
\label{app:fewshot2}

The following prompt is used in the second run to extract universal metadata related to discourse roles, citation markers, and other unequippable elements:

\begin{verbatim}
User:
Extract universal metadata from the following chunk:
Chunk: "This Agreement shall commence on the Effective Date and remain in effect..."

Assistant:
{
  "chunk": "This Agreement shall commence on the Effective Date and remain in effect...",
  "discourse_role": "introduction",
  "quotation_type": "none",
  "citation_marker": "none"
}

User:
Extract universal metadata from the following chunk:
Chunk: "The pricing for services is outlined in Exhibit A..."

Assistant:
{
  "chunk": "The pricing for services is outlined in Exhibit A...",
  "discourse_role": "body",
  "quotation_type": "none",
  "citation_marker": "exhibit_reference"
}
\end{verbatim}

\appendix
\section{Elasticsearch Schema}
\label{app:elasticsearch_schema}

The following schema is used to set up the Elasticsearch indices for both document-level and clause-level data:

\subsection{Document Index Schema}
\begin{verbatim}
PUT /document_index
{
  "mappings": {
    "properties": {
      "doc_id": { "type": "keyword" },
      "filename": { "type": "text" },
      "document_embedding": { "type": "dense_vector", "dims": 768 },
      "sections": {
        "type": "nested",
        "properties": {
          "section_title": { "type": "text" },
          "section_embedding": { "type": "dense_vector", "dims": 768 }
        }
      }
    }
  }
}
\end{verbatim}

\subsection{Clause Index Schema}
\begin{verbatim}
PUT /sentence_index
{
  "mappings": {
    "properties": {
      "doc_id": { "type": "keyword" },
      "chunk_number": { "type": "integer" },
      "chunk_text": { "type": "text" },
      "clause_type": { "type": "keyword" },
      "named_entities": { "type": "text" },
      "discourse_role": { "type": "keyword" },
      "sentence_embedding": { "type": "dense_vector", "dims": 768 },
      "re_equippable_elements": {
        "properties": {
          "quotation_marks": { "type": "boolean" },
          "citation_marker": { "type": "boolean" },
          "paraphrasing_cues": { "type": "boolean" },
          "speaker_attribution": { "type": "boolean" }
        }
      }
    }
  }
}
\end{verbatim}

\section{Embedding Generation Using the \texttt{Cohere-embed-english-v3.0} Model}
\label{app:embedding_generation}

Embeddings are generated using the \texttt{Cohere-embed-english-v3.0} model. This model produces 768-dimensional embeddings, which are indexed in Elasticsearch for vector similarity search.

\begin{itemize}
    \item \textbf{Chunk Text Embeddings}: The \texttt{chunk\_text} field is transformed into a 768-dimensional vector using the \texttt{Cohere} model.
    
    \item \textbf{Section Title Embeddings}: If present, the \texttt{section\_title} field is also embedded to provide additional context for clauses within the section.
\end{itemize}

For more information on the \texttt{Cohere-embed-english-v3.0} model, please refer to the official Cohere documentation.

\section{Few-Shot Prompting for LLM Verification}
\label{app:llm_verification_prompt}

Below is the \textit{few-shot prompt} used to guide the \textit{LLAMA 3.1 405B} model in verifying the intertextual relationships:

\begin{verbatim}
User:
Verify the relationship between these two chunks:
Chunk A: "The quick brown fox jumps over the lazy dog."
Chunk B: "The fast animal leaps over the sluggish canine."

Assistant:
{
  "relationship": "paraphrase",
  "confidence": 0.91,
  "reasoning": "Both chunks describe the same event with different wording, indicating paraphrasing."
}

User:
Verify the relationship between these two chunks:
Chunk A: "This method was used to determine X (Smith, 2020)."
Chunk B: "A similar methodology was applied (see Smith, 2020)."

Assistant:
{
  "relationship": "citation",
  "confidence": 0.95,
  "reasoning": "Both chunks cite the same method and reference the same source."
}
\end{verbatim}

This prompt helps the model classify the intertextual relationship and provides a confidence score for the result.

\end{document}